\documentstyle[preprint,tighten,aps]{revtex}

\begin{document}

\title{\bf Vacuum Instability and Pair Nucleation   \\
in a Dissipative Medium }

\author{ Roberto Iengo \\ {\it International School for Advanced Studies, Via
Beirut 4, 34014 Trieste (Italy) \\ and INFN -- Sezione di Trieste, 34100
Trieste (Italy) } }

\author{ Giancarlo Jug\cite{GJ} \\ {\it Max-Planck Institut f\"ur Physik
Komplexer Systeme, Au{\ss}enstelle Stuttgart \\
Heisenbergsrt. 1, Postfach 800665, D-70569 Stuttgart (Germany) \\
} }

\date{ \today }

\maketitle

\begin{abstract}
We present a systematic and unifying treatment of the problem of spontaneous 
nucleation of particle-antiparticle pairs in a (2+1)-dimensional system due 
to a static and uniform electromagnetic-like field, in the presence of 
quantum dissipation. We first describe a direct derivation of the 
Caldeira-Leggett type of mechanism for quantum dissipation within the 
context of string theory and of the ensuing Born-Infeld action, pointing out 
the difference with the physical context in which vacuum decay can occur.  
We then evaluate the particle-antiparticle pair production rate, working
out all the details of the calculation and including also the effects of a 
possible periodic background potential and of the Coulomb-like
particle-antiparticle attraction. The former induces a dissipation-driven 
localization which interferes with the effect of the driving electric-like
field.  We also hint at a possible application to the problem of the decay 
of a supercurrent in a superconducting thin film due to vortex-antivortex 
nucleation in the presence of a pinning lattice. 
\end{abstract}

\begin{flushleft}
PACS numbers: 11.10.-z, 11.25.-w, 05.30.-d, 74.60.Ge \\
Keywords: constant-field pair-creation, quantum dissipation 
\end{flushleft}

\section{ Introduction }
\renewcommand{\theequation}{1.\arabic{equation}}
\setcounter{equation}{0}
 
Dissipative systems are ubiquitous in nature \cite{aristotele}; 
of particular relevance in theoretical physics is ``quantum dissipation'', a 
relatively novel field of research \cite{weiss} stemming from the seminal work 
of Caldeira and Leggett (CL) \cite{cale} and which is of course of great 
relevance for low-temperature condensed-matter physics. Surprisingly, quantum 
dissipation also appears to be built-in in the dynamics of string theory 
\cite{frts,cath}.  

In this paper we systematically study the role of quantum dissipation on the 
instability of the ``vacuum'', driven by a background uniform
electromagnetic-like field. This issue is strictly related to the evaluation
of the (one-loop) effective action of a constant electromagnetic field, that
is the effective action induced by the virtual particle-antiparticle pairs. 
In fact the vacuum-instability is due in general to matter-field 
pair-nucleation and 
corresponds, technically, to the imaginary part of the effective action. In 
particular, in the context of string theory \cite{string}, there has been a 
considerable amount of work in order to determine the effective action due to 
a background electromagnetic field \cite{frts,abca}. Here we discuss and 
compare the role of quantum dissipation both in a context of string theory 
and in a context of a quantum field theory for typical condensed matter 
problems, pointing out the relevant differences and similarities.

Although parts of this paper do not contain, strictly-speaking, new results
-- since they can be found in the above-quoted string-theory papers on the
one side and in a rather condensed way in recent publications by the present 
authors \cite{ieju95,ieju96a,ieju96b} on the other side -- we do think it is 
important to offer a systematic and unified treatment of quantum dissipation 
and vacuum decay which can be of interest to both the string-theory and the 
condensed-matter physics communities. Furthermore, we do believe that the 
blend of many ingredients as presented in this paper is novel, timely and 
compelling. 

The organization of this article is indeed as follows. We first (Section II)
discuss quantum dissipation in the context of string theory, and exhibit a
direct computation which shows that the string degrees of freedom provide an 
ideal CL dissipative bath for the string's end-point. In Section III, we 
complete the discussion pertinent to string theory by showing how quantum
dissipation in a totally relativistic context yields the so-called Born-Infeld
effective action \cite{boin} for the electromagnetic field. Namely, in this
context, dissipation affects the motion in the time- and in the 
space-coordinates on the same footing. In this case we find that there is no 
vacuum decay. However, in a field-theory context more 
specifically oriented towards condensed-matter situations, where quantum
dissipation is originated by the interaction with a medium which has only 
space reality, the vacuum can indeed decay and the calculation for the 
associated rate of pair production is described in Section IV. There, we 
specialize the calculation for the case of two space dimensions, by
recalling the relevance of this calculation for the vortex-nucleation in a
thin superconducting film. In Section V we introduce a physics problem where 
quantum dissipation plays an essential role, namely the dissipation-driven
localization transition in the presence of a periodic background potential.
Finally, in Section VI, we describe the vacuum instability when both 
background uniform electromagnetic-like field and a periodic background 
potential are present. We also include the effect of a Coulomb-like attraction
between the particle and antiparticle in a pair. In Section VII we present 
our conclusions, commenting on the applicability of our work to the problem 
of vortex-antivortex nucleation-induced decay of a supercurrent in a 
superconducting thin film where a periodic pinning lattice has been 
artificially introduced.
     
We point out from the outset that since the phenomenon of spontaneous 
pair-nucleation driven by a uniform and static field is a long-standing 
prediction of quantum field theory which has never been directly observed so 
far -- to our knowledge -- our work on quantum vortex-nucleation could 
represent the possibility for an experimental observation of this phenomenon. 
Since this observation would occur in a solid-state device, the role of 
quantum dissipation is compelling.

\section{ Quantum Dissipation in String Theory }
\renewcommand{\theequation}{2.\arabic{equation}}
\setcounter{equation}{0}

The results reported in this Section and in the following one can also be
found in the papers by Fradkin and Tseytlin \cite{frts}, by Abouelsaood, 
Callan, Nappi and Yost \cite{abca}, and by Callan and Thorlacius \cite{cath}. 
However here we present an alternative derivation, making use of a direct 
and straightforward computation which has also the advantage of being easily 
compared with the standard Caldeira and Leggett formulation \cite{cale} and 
with the other physical contexts to be discussed in the rest of this paper.  
We begin by writing, quite generally, the (Euclidean) action for an open 
string

\begin{equation}
{\cal S}_E=\int_0^{\tau}dt\int_0^{\pi}d\sigma \left ( \frac{\rho}{2}
\left ( \frac{\partial X}{\partial t} \right )^2 +\frac{\nu}{2} \left ( 
\frac{\partial X}{\partial \sigma} \right )^2 \right ).
\end{equation}

\noindent
Here, $X=X_{\mu}(t,\sigma)$ are the space-time coordinates of the string,  
whereas $t$ plays the role of the Schwinger (Euclidean) proper time
parameter and $\sigma$ parametrizes the extension of the string.
We consider periodic boundary conditions in $t$, and Neumann boundary
conditions at the end points in $\sigma$: 
$\partial_{\sigma}X|_{\sigma=0}=\partial_{\sigma}X|_{\sigma=\pi}=0$.
The manifold spanned by the string during its motion in $t$ is thus an 
annulus. We have introduced two arbitrary parameters for generality's sake,
$\rho$ playing the role of the mass density and $\nu$ that of an elastic
constant. In the standard string theory \cite{string}, they are related to 
each other and to the string tension. 

We now derive the dynamics of the end point of this string, say at 
$\sigma=0$, denoting $q(t){\equiv}X(t,\sigma=0)$, by deriving the end 
point's effective action, $S(q)$. We begin by Fourier analysing in $\sigma$

\begin{equation}
X(t,\sigma)=\sum_{n=0}^{\infty} X_n(t) \cos (n\sigma),
\end{equation}

\noindent
and further we Fourier analyse in $t$: 
$X_n(t)=\sum_{k=-\infty}^{\infty} X_{nk} e^{ i\frac{2\pi k}{\tau} t }$. 
The Euclidean action is now

\begin{equation}
{\cal S}_E=\frac{\pi}{4}\tau\nu \sum_{n=0} n^2 X_{n0}^2
+ \frac{\pi}{2}\tau \sum_{n=0}\sum_{k=1} \left ( \rho 
\left ( \frac{2\pi k}{\tau} \right )^2 + n^2\nu \right ) |X_{nk}|^2, 
\end{equation}

\noindent
and the effective end-point action is obtained by tracing out all string
degrees of freedom, except those for the end point. We thus compute the 
constrained functional integral 

\begin{equation}
e^{-S(q)}={\cal N} \int \prod_{n=0} \left \{ dX_{n0} 
\prod_{k=1} d^2X_{nk}  \right \} 
e^{-{\cal S}_E} ~ \delta(\sum_{n=0}X_{n0}-q_0) \prod_{k=1}
\delta(\sum_{n=0}X_{nk}-q_k),
\end{equation}

\noindent
having introduced 
$q(t)=\sum_{k=-\infty}^{+\infty} q_k e^{ i\frac{2\pi k}{\tau} t }$.
Now for each $k$ we evaluate the constrained functional integral, by
replacing $X_{0k}=q_k-\sum_{n=1}X_{nk}$. We have

\begin{equation}
e^{-S(q_k)}=e^{ -\frac{\pi}{2}\rho\tau \left ( \frac{2\pi k}{\tau} \right )^2
|q_k|^2 } {\cal N} \int \prod_{n=1} d^2X_{nk} \exp \{ -X^{*}_{nk}B_{nl}X_{lk}
-X^{*}_{nk}V_n-V^{*}_nX_{nk} \},
\end{equation}

\noindent
where we have introduced the matrix ($n,l=1,2,\dots,\infty $)

\begin{equation}
B_{nl}=a(n)\delta_{nl} +a(0)
\end{equation}

\noindent
and the vector $V_n=a(0) q_k$, which is actually independent of $n$. Here

\begin{equation}
a(n)=\frac{\pi}{2}\tau \left ( \rho \left ( \frac{2\pi k}{\tau} \right )^2
+ n^2\nu \right ).
\end{equation} 

\noindent
Completing the square, we find

\begin{equation}
e^{-S(q_k)}=e^{ -\frac{\pi}{2}\rho\tau \left ( \frac{2\pi k}{\tau} \right )^2
|q_k|^2 } \exp \{ V^{*}_n B^{-1}_{nl} V_l \}.
\end{equation}

\noindent
One can check that the inverse matrix is

\begin{equation}
B^{-1}_{nl}=\frac{1}{a(n)}\delta_{nl}-\frac{a(0)}{a(n)Qa(l)},
\end{equation}

\noindent
where we have set $Q=\sum_{n=0}\frac{a(0)}{a(n)}$. We get, therefore

\begin{equation}
S(q_k)=\frac{1}{ \sum_{n=0} 1/a(n) } |q_k|^2.
\end{equation}

\noindent
Going now to the continuum limit in $\sigma$, which corresponds to the case
where the inner circular border of the annulus shrinks to zero and 
mathematically to the limit $\nu{\rightarrow}0$ (but keeping $\rho\nu$ 
finite), we replace the sum over $n$ with an integral, to get (regardless  
of the sign of $k$)

\begin{equation}
S(q_k)=\frac{ \pi\nu\tau/2 }{ \int_0^{\infty} dx/ \left ( \frac{\rho}{\nu}
\left ( \frac{2\pi k}{\tau} \right )^2 +x^2 \right ) } |q_k|^2=
\eta 2\pi |k| |q_k|^2.
\end{equation}

\noindent
Here $\eta=\sqrt{\rho\nu}$ is to be identified with the 
friction coefficient of the CL formulation of quantum dissipation 
\cite{cale}. Interestingly, we obtain pure dissipative dynamics for the 
end-point. By Fourier transformation of this expression we get the standard 
form for the periodic case

\begin{equation}
S(q)
=\frac{\eta\pi}{4\tau^2} \int_0^{\tau} dt \int_0^{\tau} dt'
\left ( \frac{ q(t)- q(t') }{ \sin [ \pi(t-t')/\tau ] } 
\right )^2.
\label{fulldiss}
\end{equation}

\noindent
Notice that the true CL dissipative action \cite{cale} is obtained only 
when the string coordinates are strictly space-like.

\section{ Effective Electromagnetic Action in String Theory }
\renewcommand{\theequation}{3.\arabic{equation}}
\setcounter{equation}{0}
 
Since $X(t,\sigma)$, and thus also $q(t)$, represent the complete set of
space-time coordinates, the effective action for the end point found in the
previous Section II contains quantum dissipation also in the time 
coordinate, $q_{\rm time}(t)$. In this situation it is immediate to work out 
the effective action pertinent to the situation where a uniform 
electromagnetic field is coupled only to the string's end point. Namely, let 
us write the full action for $q(t)$, including this coupling

\begin{eqnarray}
S(q,A_{\mu})&=&\sum_{k=1}^{\infty} \left [ \eta 2\pi k |q_{\mu k}|^2
+ i F_{\mu\nu} (i 2\pi k) q_{\mu k}^{*} q_{\nu k} \right ] \\ \nonumber
&=&
\sum_{k=1}^{\infty} 2\pi k q_{\mu k}^{*} \left \{ \eta\delta_{\mu\nu}
- F_{\mu\nu} \right \} q_{\nu k}.
\end{eqnarray}

\noindent
Therefore, we get the effective action $S_{eff}(A_{\mu})$ (remembering that
the result of the functional integration over the string coordinates $X$
is interpreted in string theory as the one-loop contribution to the 
effective action $S_{eff}$):

\begin{equation}
S_{eff}(A_{\mu})={\cal N} \int {\cal D}q ~ e^{ - S(q,A_{\mu}) } =
\prod_{k=1}^{\infty} det \left ( \delta_{\mu\nu}-\frac{1}{\eta} F_{\mu\nu}
\right )^{-1} = \sqrt{ det \left ( \delta_{\mu\nu}-\frac{1}{\eta} F_{\mu\nu}
\right ) }.
\end{equation}

\noindent
We have used here the standard $\zeta$-function regularization. We have thus
obtained the Born-Infeld action \cite{boin} in a rather straightforward way. 
We notice that this action is real (at least for fields such that, 
generically speaking, $F < \eta$) and there is thus no vacuum decay. 

\section{ Vacuum Instability in a Dissipative Medium }
\renewcommand{\theequation}{4.\arabic{equation}}
\setcounter{equation}{0}

In the following, we discuss situations where dissipation does lead to a 
vacuum decay, that is when it appears only for the space coordinates. This 
is indeed the physical situation for a particle which during its space-time 
motion interacts with the harmonic oscillators of a suitable CL bath. The 
particle will feel interactions which will depend on its space position and 
will affect its space trajectory. In these situations the action for the 
time-like component of the coordinates is of the usual form, namely
$\int_0^{\tau} dt \frac{1}{4} ( \partial_t q_{\rm time}(t) )^2$. 

We consider here the space-time coordinate $q(t)$ to describe closed loops 
in space-time as a function of the Schwinger (Euclidean) proper time $t$. 
These will correspond to the vacuum fluctuations due to virtual creation and 
annihilation of particle-antiparticle pairs. Vacuum instability will manifest
itself when the virtual loops extend to infinity, giving rise to an imaginary
part for the vacuum energy $W_0$. The latter, before the introduction of
dissipation, is by definition ($T$ being the total time interval)

\begin{equation} 
W_0= \frac{1}{T} Tr \ln \left (
-\frac{1}{\gamma}{\bf D}^2-D_{\rm time}^2+{\cal E}_0^2 \right ).
\label{vacuum}
\end{equation}

\noindent
This equation represents the standard contribution to the vacuum energy of a
complex scalar field, whose quanta constitute the particles-antiparticles. 
$D$ is here the covariant derivative, containing the electromagnetic field
$A_{\mu}$, ${\cal E}_0$ is the activation energy -- that is, in the 
relativistic field theory, the rest mass $m$ times $c^2$, the square 
velocity of light -- and $\gamma=1/c^2=m/{\cal E}_0$. In turn, we write

\begin{equation} 
Tr \ln \frac{-D_E^2+{\cal
E}_0^2}{\Lambda^2}=-\lim_{{\epsilon}{\rightarrow}0}\int_{\epsilon}^{\infty}
\frac{d\tau}{\tau} Tr \left \{ e^{-(-D_E^2+{\cal E}_0^2)\tau} -
e^{-{\Lambda}^2\tau} \right \}, 
\label{identity} 
\end{equation}
   
\noindent
where we have defined 
$D^2_E=\frac{1}{\gamma}{\bf D}^2+D_{\rm time}^2$. Finally, we use the
path-integral representation (evaluating the imaginary part $\Gamma/2$ of 
the vacuum energy)

\begin{equation} 
\frac{\Gamma}{2}=\frac{1}{T} ~ Im ~ \int_0^{\infty} \frac{d\tau}{\tau} 
{\cal N} \int_{q(0)=q(\tau)} {\cal D}q(t) e^{ -S_E(q,A_{\mu})}.
\label{masterrate} 
\end{equation}
 
\noindent
Quantum dissipation is now introduced by writing the corresponding Euclidean
action, that is (restricting the form Eq. (\ref{fulldiss}) to the sole
spatial coordinates)

\begin{eqnarray}
S_E(q,A_{\mu})&=&\int_0^{\tau} dt \left \{ 
\frac{1}{4} ( \partial_t q_{\rm time} )^2 + \frac{\gamma}{4} 
( \partial_t {\bf q} )^2 +i \partial_t q_{\mu} A_{\mu}(q) 
+ {\cal E}_0^2 \right \} \\ \nonumber 
&+&\frac{\eta\pi}{4\tau^2} \int_0^{\tau} dt \int_0^{\tau} dt'
\left ( \frac{ {\bf q}(t)-{\bf q}(t') }{ \sin [ \pi(t-t')/\tau ] } 
\right )^2,
\label{action}
\end{eqnarray}

\noindent
which in terms of the Fourier components of $q(t)$ reads

\begin{equation}
S_E(q,A_{\mu})=\sum_{k=1}^{\infty} \left \{ \frac{1}{2} 
\frac{(2\pi k)^2}{\tau} |q_{{\rm time} ~ k}|^2 + \left [ \frac{\gamma}{2}
\frac{(2\pi k)^2}{\tau} + \eta 2\pi k \right ] |{\bf q}_k|^2 
-F_{\mu\nu} (2\pi k) q_{\mu k}^{*}q_{\nu k} \right \} 
+{\cal E}_0^2\tau.
\end{equation}

\noindent
We are interested in the dissipation-dominated case, corresponding to 
$\gamma{\rightarrow}0$. We make use of a $\gamma{\neq}0$ as a regulator for
the intermediate calculations. 

>From here on, we concentrate on the case of two space dimensions, having in 
mind the possible application to vortex-antivortex pair nucleation in
superconducting thin films (Section VII). The electromagnetic field is taken 
to have the electric components ${\bf E}=(E,0)$ in the two-dimensional 
plane, and the magnetic component $B$ orthogonal to the plane. As it is 
known, and as was shown in Ref. \cite{ieju95}, the effect of the magnetic 
part $B$ effectively amounts to a renormalization of the friction coefficient:
$\eta{\rightarrow}(B^2+\eta^2)/\eta$. Therefore, in what follows we drop
$B$ and simply denote by $\eta$ the renormalized dissipation coefficient.   
Notice that the electric field appears as an imaginary quantity in the 
Euclidean formulation, that is $F_{a ~ {\rm time}}=(iE,0)$ with $a=x, y$.

As a first step in the computation, we perform the integral over 
$q_{{\rm time}}$. We get, by shifting the integration variable and performing
the Gaussian integration: 

\begin{eqnarray}
&&{\cal N}
\int_{q(0)=q(\tau)} {\cal D}q(t) ~ e^{ -S_E(q,A_{\mu}) }= \\ \nonumber
&&=\left ( \frac{1}{4\pi\tau} \right )^{1/2} e^{-{\cal E}_0^2\tau} ~
{\cal N} \prod_{k=1}^{\infty} \int d^2q_{xk} e^{ -( \eta 2\pi k
-2\tau E^2 ) |q_{xk}|^2 } \int d^2q_{yk}  e^{ -\eta 2\pi k |q_{yk}|^2 }.
\end{eqnarray}

\noindent
The factor $\left ( \frac{1}{4\pi\tau} \right )^{1/2}$ is here the standard
free-particle result for the path-integral over $q_{\rm time}$.
Now we come to the essential point of our computation. The integration over
$q_{xk}$ gives rise to poles in the variable $\tau$, which in turn -- when
inserted in the integration over $\tau$ with a suitable $i\epsilon$
prescription ($Im \frac{1}{x-i\epsilon}=\pi\delta(x)$) -- give rise to an 
imaginary part for the vacuum energy, Eq. (\ref{vacuum}). The dominant pole, 
corresponding to the minimal value of $\tau$ and thus giving the leading 
term in an expansion made up of further exponentially-suppressed 
contributions, comes from integrating over $q_{x1}$:

\begin{equation}
\int d^2q_{x1}e^{ -( \eta 2\pi - 2\tau E^2 ) |q_{x1}|^2 }=
\pi/ \left ( \eta 2\pi - 2\tau E^2 \right ).
\label{dominant}
\end{equation}

\noindent
Thus we keep only the dominant pole $\tau=\tau_1=\pi\eta/E^2$. We are left
with the task of computing the ``entropy'' due to the integration over the
remaining $q_{xk}$, $q_{yk}$. On the pole, the action for the modes 
$q_{x ~ k>1}$ is the same as for $E^2=0$, as it is for $q_{yk}$, but with 
$k{\rightarrow}k-1$. Thus we compute (for a one-dimensional case)  

\begin{equation}
Z_0{\equiv}{\cal N} \prod_{k=1} \int d^2q_k e^{ - \eta 2\pi k |q_k|^2 }       
={\cal N}\prod_{k=1} \frac{1}{ 2\eta k }. 
\label{entropy}
\end{equation}

\noindent
Bearing in mind the normalization for the free particle,
${\cal N}\prod_{k=1} \frac{\tau}{ 2\pi\gamma k^2 }=
\left ( \frac{\gamma}{4\pi\tau} \right )^{1/2}$, we regularize the above 
expression by introducing a frequency cutoff $\Omega$ for the dissipation 
as well as reintroducing the inertial mass represented by $\gamma$. Thus 
we write

\begin{equation}
{\cal N}\prod_{k=1} \frac{1}{ 2\eta k }{\rightarrow}
\left ( \frac{\gamma}{4\pi\tau} \right )^{1/2} \prod_{k=1}^{n^{*}} 
\left ( 1+\frac{\tau\eta}{\gamma\pi k} \right )^{-1},
\label{cutprod} 
\end{equation}

\noindent
where $n^{*}=\Omega\tau_1/2\pi$ is a large integer. Therefore, by using the
formula 

\begin{equation}
\prod_{k=1}^{n^{*}} \left ( 1+\frac{z}{k} \right )^{-1}{\rightarrow} ~
(n^{*})^{-z} \Gamma(z+1),
\end{equation}

\noindent
and the large-argument expression for the $\Gamma$-function, we find

\begin{equation}
Z_0(x)Z_0(y)=\frac{\eta}{2\pi} e^{-\tau_1\Delta{\cal E}^2}.
\label{xyentropy}
\end{equation}

\noindent
Thus we see that the regularization cutoffs $\Omega$ and $\gamma$ are
reabsorbed into a renormalization of the activation energy:

\begin{equation}
\Delta{\cal E}^2=
\frac{2\eta}{\pi\gamma} (1+\ln \frac{\Omega\gamma}{2\eta}). 
\end{equation}

\noindent
Here, we will include $\Delta{\cal E}^2$ into a redefinition of the 
activation energy,
${\cal E}_0^2{\rightarrow} {\cal E}_R^2={\cal E}_0^2+\Delta{\cal E}^2$,
so that, effectively, $Z_0(x)Z_0(y)=\eta/2\pi$. Actually, an extra factor of
$2\eta$ is retrieved from the fact that in the zero-field case (used to fix 
the normalization) the $q_{x1}$-mode contribution is assumed to be included.
Since here we instead compute its contribution separately, we have to divide
by $\int d^2q_{x1} e^{-2\pi\eta |q_{x1}|^2}=1/2\eta$.

Finally, by carrying out the imaginary part of the $\tau$-integral in
Eq. (\ref{masterrate}), and factoring out the integration over the zero 
modes $q_0$, giving the space-time volume, we get the pair-production rate 
per unit area

\begin{equation}
\frac{\Gamma}{2L^2}=\frac{E\sqrt{\eta}}{2\pi} 
e^{-\frac{\pi\eta{\cal E}_R^2}{E^2}}.
\label{rate0}
\end{equation}

\noindent
This formula applies to vortex-nucleation in a superconducting thin film
(\cite{ieju95}, Section VII), when we set $E=2\pi J$, $J$ being the 
supercurrent number density and ${\cal E}_R$ a suitably renormalized 
nucleation energy.
The remainder of the paper deals with the interesting problem of vacuum 
decay in a dissipative medium when a background periodic potential is also
present. 

\section{ Dissipation-Driven Localization Transition }
\renewcommand{\theequation}{5.\arabic{equation}}
\setcounter{equation}{0}

This Section summarises, for the reader's convenience, some known facts about
the problem of the diffusion of a quantum particle in a dissipative medium, in
the presence of a periodic potential. In the next Section, we will adapt this
discussion for the case of pair-nucleation in the presence of a background
periodic potential, of the form

\begin{equation}
V({\bf q})=A_0 \sum_a \left (1-\cos kq_a \right ),
\label{potent}
\end{equation} 

\noindent
where the sum extends over the space coordinates. As is well-known 
\cite{schmid,ghm}, the motion of a quantum particle in a periodic potential 
in the presence of friction displays a localization transition whereby for 
$\eta < \eta_c$ the particle is mobile whilst for $\eta > \eta_c$ it is 
confined, $\eta_c=k^2/2\pi=2\pi/d^2$ being a sharp threshold independent of 
the potential's amplitude $A_0$ (here, $d=2\pi/k$ is the potential's lattice
parameter; we assume a square-lattice structure). This is reproduced by our 
formalism, with ${\bf E}=B=0$ and $\gamma{\rightarrow}0$ in 
Eq. (\ref{action}), with the confined phase treated within the variational 
``blocked-renormalization'' scheme first introduced by Fisher and Zwerger
\cite{fizw}, through which a ``mass'' is dynamically generated for the 
damped modes' propagator. We recall that the presence of the potential 
$V({\bf  q})$, Eq. (\ref{potent}), modifies the action by a term 
$\int_0^{\tau} dt V({\bf  q})$. In the next Section we will come
back to the case ${\bf E}\not= 0$ to see how the
presence of the periodic potential affects the pair production
(the case ${\bf E}= 0$ and $B\not= 0$ has been discussed in Ref.
\cite{cafe}).  

It is actually useful to recall here some of the results of this 
Kosterlitz-like renormalization group (RG) analysis, as it will be of some 
use in our treatment for the nucleation rate (Section VI). In our notation, 
the motion of a quantum particle moving in a dissipative environment and 
subjected to a periodic potential (\ref{potent}) is described by the 
Euclidean action

\begin{equation}
{\cal H}({\bf q})
=\sum_{n=1} \eta\tau\omega_n q_a(\omega_n)q_a^{*}(\omega_n)
+\int_0^{\tau} dt ~ A_0\sum_a \left ( 1- \cos kq_a \right ),
\label{hdiff}
\end{equation}

\noindent
where $\omega_n=\frac{2\pi}{\tau}n$, and $q(\omega_n)$ are the Fourier 
modes $q_n$. In the continuum limit, the bare propagator is 

\begin{equation}
\langle q_a(\omega)q_{a'}^{*}(\omega ') \rangle_0=
\delta (\omega-\omega ') \delta_{aa'}\frac{1}{\eta |\omega|}
f \left ( \frac{|\omega|}{\Omega} \right ),
\label{propagator}
\end{equation}

\noindent
with $f(x)$ a cutoff function and $\Omega$ the same CL cutoff frequency
as in Section IV. We can, for example, make use of a frequency-space
RG procedure at this point, much as in the theory of the surface
roughening transition \cite{nozieres}. Writing 
$q_a(t)=\hat{q}_a(t)+\delta q_a(t)$, where $\delta q_a(t)$ contains the
modes with frequencies $\frac{\Omega}{b} < \omega < \Omega$ ($b=1+\epsilon$,
with $\epsilon\rightarrow 0$), and integrating out these modes one finds 
that the first-order contribution represents a renormalization 
of the amplitude $A_0$ in the action. For this we get: 

\begin{equation}
\hat{A}_0=A_0\exp\{-\frac{1}{2}k^2 \langle (\delta q_a)^2 \rangle_0 \}
=A_0e^{-\frac{1}{2\pi\eta}k^2\ln(1+\epsilon)}.
\end{equation}

\noindent
There is no renormalization of the friction coefficient, and this can be
argued \cite{fizw} to hold to all orders. The RG recursion 
relations are then, in differential form (with $d\ell=\ln (1+\epsilon)$):

\begin{eqnarray}
\frac{dA_0}{d\ell}&=& - \frac{k^2}{2\pi\eta}  
A_0+\cdots = - \frac{\eta_c}{\eta} A_0+\cdots \nonumber \\
\frac{d\eta}{d\ell}&=&0.
\label{rgrec}
\end{eqnarray}

\noindent
To solve these is straightforward in the mobile phase, 
$\eta < \eta_c=k^2/2\pi$:

\begin{equation}
A_{0R}(\ell)=A_0(0)\exp\{-\frac{\eta_c}{\eta}\ell \}, 
\qquad \eta(\ell)=\eta(0),
\end{equation}

\noindent
when the dimensionless amplitude of the renormalized pinning potential, 
${\cal A}_{0R}(\ell)=A_{0R}(\ell)/\Omega(\ell)$, vanishes exponentially for
increasing time-scales (here $\Omega(\ell)=\Omega e^{-\ell}$).
However, in the confined phase, $\eta > \eta_c$,
the amplitude diverges, the perturbation expansion in $A_0$ breaks
down and the RG iteration must be interrupted. Since the non-renormalization 
of $\eta$ allows for no natural characteristic time-scale (unlike in a
genuine Kosterlitz-Thouless transition) to be introduced, a 
dynamically-generated mass term is included in the free Hamiltonian 
\cite{fizw}. Carrying out the RG procedure with a modified propagator 

\begin{equation}
\langle q_a(\omega)q_{a'}^{*}(\omega') \rangle_0=
\delta (\omega-\omega ') \delta_{aa'}\frac{1}{\eta |\omega|+M^2}
f \left ( \frac{|\omega|}{\Omega} \right ),
\label{mpropagator}
\end{equation}

\noindent
one arrives at the differential RG equation for $A_0$:

\begin{equation}
\frac{dA_0}{d\ell}=-\frac{\eta_c}{\eta}\frac{1}{1+\mu e^{\ell}}A_0+\cdots,
\end{equation}

\noindent
where $\mu$ is the dimensionless square ``mass'' $\mu=M^2/\eta\Omega$, which
has the solution (assuming $\mu$ scale-independent):

\begin{equation}
A_{0R}(\ell)=A_0 \exp \left \{ -\frac{\eta_c}{\eta}\ln \left ( 
\frac{1+\mu}{e^{-\ell}+\mu} \right ) \right \},
\label{flowampl}
\end{equation}

\noindent
corresponding to the change of pinning potential amplitude following an
integration of the degrees of freedom in the shell 
$\Omega(\ell) < \omega < \Omega$. 
The mass is then determined by a self-consistent procedure (Feynman's 
variational principle) \cite{fizw} minimising the system's free energy with 
respect to $\mu$, a procedure leading to

\begin{equation}
\mu=\frac{A_0k^2}{\eta\Omega}(\mu)^{\eta_c/\eta}.
\end{equation}    

\noindent
This equation has the solution $\mu=0$ in the mobile phase ($\eta < \eta_c$)
and, in the more interesting confined phase ($\eta > \eta_c$):

\begin{equation}
\mu=\left ( \frac{2\pi A_0}{\Omega} \frac{\eta_c}{\eta} \right )^{\eta/
(\eta-\eta_c)}.
\end{equation}

\noindent
This concludes the necessary repilogue of the dissipation-driven localization
transition.

\section{ Pair-Production Rate in the Presence of a Periodic Potential }
\renewcommand{\theequation}{6.\arabic{equation}}
\setcounter{equation}{0}

We now come to the evaluation of $\Gamma$, in the presence of the periodic
potential of Eq. (\ref{potent}), which calls for the implementation of a 
RG-type approach. For reasons related to the physics of vortices in a 
superconductor, we are in fact interested in the case of strong dissipation 
(confined phase, $\eta > \eta_c$). In the relevant case and with the 
electric-field ${\bf E}$ in the $x$-direction, the motion of the quantum 
particle in the orthogonal $y$-direction is as described in the previous  
Section V and with the dynamically-generated ``mass'' treatment.
Along the $x$-direction, we have seen that the mode $q_1$ is driven to
infinity, while all other $n > 1$ modes remain bounded. We can regard the
modes with $n > 1$ as describing fluctuations around the mean trajectory of
the nucleating pair. Therefore, with the notation of the previous Section, 
we take $\hat{q}(t)=q_0+q_1e^{i\omega_1 t}+q_1^{*}e^{- i\omega_1 t}$ as the
RG survivor-modes and $q_R(t)$ as the remainder, RG-integrated coordinate 
modes which infinitesimal part is the $\delta q(t)$ introduced before. In 
order to take into account the effect of the periodic potential in the 
$x$-direction we assign the dynamically-generated mass also to $q_R$. Now, 
however, we must take into account that the amplitude of the potential as
``seen'' by $q_R$ still depends on $\hat{q}$. Therefore, the 
variationally-determined mass will be in the end evaluated self-consistently. 

The effect of the dynamically-generated mass on the computation will be
encoded in the renormalization of $A_0$ and  in the evaluation of the 
``entropy'' factor $Z$ due to the $n > 1$ modes in the $x$-direction ($q_R$), 
and to all $n > 0$ modes in the $y$-direction. Since the modes in $q_R$ are 
purely Gaussian, we can now evaluate (for each space dimension) the 
correcting factor $B^{*}$ due to the dynamically-generated mass $M$: 
$Z=Z_0B^{*}$, where $Z_0$ is given by Eq.s (\ref{entropy}) and 
(\ref{xyentropy}) and where

\begin{equation}
B^{*}=\prod_{n=1}^{n^{*}} \left ( 1+\frac{\tau M^2}{2\pi\eta n} \right )^{-1}.
\end{equation}

\noindent
Making use of arguments similar to those seen for Eq.s (\ref{cutprod}) and
(\ref{xyentropy}) we get (in the limit where the pole value $\tau_1$ is very 
large)

\begin{equation}
B^{*}=\sqrt{ \frac{M^2\tau_1}{\eta} } e^{ -\tau_1\Delta\tilde{\cal E}^2 },
\end{equation}

\noindent
where 
$\Delta\tilde{\cal E}^2=\frac{M^2}{2\pi\eta}
(1+\ln \frac{\Omega\eta}{M^2})$
is to be regarded as a further renormalization of the activation energy.

Having treated the effect of the higher modes in $q_R$, we now discuss the
dramatic dynamical effects of the unbounded $\hat{q}(t)$. The dynamics of
these surviving modes is described by the Euclidean action $\hat{S}_E$:

\begin{equation}  
\hat{S}_E=( 2\pi\eta - 2\tau E^2 )|q_1|^2 + A_{0R}\int_0^{\tau} dt ~ 
( 1- \cos k\hat{q}(t) ) + \tau \left ( {\cal E}^2_R + K \ln \frac{|q_1|}{a} 
\right )
\end{equation}

\noindent
In this formula we have also introduced, for completeness, the Coulomb 
interaction between the particle and the antiparticle of the pair, in two 
space dimensions, $a$ being a suitable ultraviolet cutoff. We treat this 
Coulomb interaction approximately, keeping only the dependence on the 
average particle-antiparticle distance, $|q_1|$. $A_{0R}$ is given by 
Eq. (\ref{flowampl}), with $\ell=\ell^{*}=\ln n^{*}{\rightarrow}\infty$.
We find ($J_0(x)$ and $I_0(x)$ being Bessel functions)

\begin{equation}
\int_0^{\tau} \cos k\hat{q}(t)=\tau J_0( 2k |q_1| )\cos kq_0, 
\end{equation}

\noindent
and, averaging over the overall position $q_0$ 

\begin{equation}
\hat{S}_E=( 2\pi\eta - 2\tau E^2 )|q_1|^2 - \ln \left ( I_0 \left ( \tau 
A_{0R}J_0 ( 2k |q_1| ) \right ) \right )
 + \tau \left [ {\cal E}^2_R + K \ln \frac{|q_1|}{a} + A_{0R} \right ].
\end{equation}

\noindent
The integration $\int d^2q_1 e^{-\hat{S}_E}$ and the subsequent integration
over $\tau$, as in Eq.s (\ref{masterrate}) and (\ref{dominant}), can then 
be performed by means of a saddle-point method, making use of the asymptotic 
formula for large argument    

\begin{equation}
I_0(x\tau){\rightarrow}\frac{1}{\sqrt{2\pi |x|\tau}} e^{ |x|\tau }.
\end{equation}

\noindent
Equating to zero the variation of $\hat{S}_E$ with respect to $\tau$ gives 
implicitely the value of $\ell_N{\equiv}|q_1|$ at the saddle point

\begin{equation}
\ell_N^2=\frac{1}{2E^2} \left (
 {\cal E}^2_R + K \ln \frac{\ell_N}{a} 
 + A_{0R} - |A_{0R}J_0( 2k \ell_N )|  \right ).
\end{equation}

\noindent
This equation can be solved recursively. At the leading order we
find the same solution of Section V, namely 
$\ell_N^2=\frac{1}{2E^2} \left (
 {\cal E^{'}}_R^2 + A_{0R}^{'} \right )$, now having set
${\cal E^{'}}^2_R={\cal E}^2_R+K \ln (\ell_{0N} /a)$ and 
$A^{'}_{0R}=A_{0R}-|A_{0R}J_0(2k\ell_{0N} )|$ with
$\ell_{0N}=({\cal E}^2_R+A_{0R})/2E^2$. 

Similarly, the leading saddle-point value for $\tau$ is the same as in 
Section IV, $\tau_1=\pi\eta/E^2$. Thus, at the leading order the $q_1$ and
$\tau$ integrals can be done as in Section IV. Therefore, we get   

\begin{equation}
\frac{\Gamma}{2L^2}=B^{*}_xB^{*}_y 
\frac{E\sqrt{\eta}}{2\pi} I_0 \left ( \frac{\pi\eta A_{0R}}{E^2} J_0(2k\ell_N)
\right ) 
\exp \{-\frac{\pi\eta \left ( {\cal E}^2_R+K\ln(\ell_N/a)+A_{0R} 
\right ) }{E^2} \}. 
\end{equation}

\noindent
It is important to notice that, strictly speaking, the dynamically-generated
``masses'' are different in the $x$- and $y$-directions. Moreover, we have yet
to determine the self-consistent value of the mass for the $x$-direction; 
the latter can be carried out by repeating the variational calculation 
reviewed in Section V and by substituting for $A_0$: 
$A_0{\rightarrow}A_0 |J_0(2k\ell_N)|$. This, of course, holds good only 
away from the
zeroes of the $J_0(x)$-function where also the whole saddle-point evaluation
above should be modified. Still, these zeroes denote minima of the production
rate (the latter being our main physical prediction here) where the 
pair-nucleation has a deep drop. This is the main physical effect, and thus 
details of the dynamics inside these minima are believed to be irrelevant. 

Our final formula for the production rate, neglecting for simplicity's sake
possible differences between the $x$- and $y$- dynamically-generated masses,
and using the dimensionless quantity $\mu=M^2/\Omega\eta$, is:

\begin{equation}
\frac{\Gamma}{2L^2}= 
\frac{\mu\Omega}{2E}\eta^{3/2} 
I_0 \left ( \frac{\pi\eta A_{0R}}{E^2} J_0(2k\ell_N) \right )
\exp \{ -\frac{\pi\eta \left ( {\cal E}^2_R+K\ln(\ell_N/a)+A_{0R} 
\right ) }{E^2} \}.
\label{fullrate}
\end{equation}

\section{ Possible Application to Vortex-Nucleation Induced Decay
of a Supercurrent in a Superconducting Thin-Film. }
\renewcommand{\theequation}{7.\arabic{equation}}
\setcounter{equation}{0}
 
In some recent papers \cite{ieju95,ieju96a}, we have argued that our theory 
could find its most direct application (that is, beyond its intrinsic 
theoretical physics value) in a specific solid-state physics problem. 
Namely, we have considered the situation in which -- in the absence of an 
external magnetic field -- a supercurrent is maintained in a thin 
superconducting film at very low temperatures, where the vortex dynamics is 
intrinsically quantum. We envisage films made up of some high-$T_c$ material, 
in order to afford high critical currents. Experiments \cite{paro} have 
shown evidence for a residual low-temperature resistance which depends very 
non-linearly on the current, $I$. We have argued that, beside the standard 
edge-tunneling mechanism \cite{aoth93,stephen,ieju96b}, spontaneous 
vortex-antivortex nucleation should also play an important role in inducing 
supercurrent decay (see also Ref. \cite{ao} for earlier work).

We very briefly recall how the problems discussed in the previous Sections can
find their implementation in vortex physics \cite{reviews}. Vortices in a
superconducting material experience a Magnus force \cite{reviews,aoth94}
(even in the absence of an external magnetic field) which is the exact 
analogue of the electromagnetic force for point-charges and couples to the
vortices via their topological charge. One can show that while the 
standard expression for the 
magnetic-like Magnus field, orthogonal to the film's plane in our geometry, 
is given by $B_M=2\pi s\rho^{(3)}_s$ ($s$ being the film's thickness, 
$\rho^{(3)}_s$ the three-dimensional superfluid number density), the 
electric-like part is proportional and orthogonal to the applied supercurrent 
number density ${\bf J}$, that is ${\bf E}_M=\times 2\pi {\bf J}$ 
\cite{ieju95}. The Magnus force is then of the Lorentz form, namely
${\bf F}_M= {\bf E}_M-\dot{\bf q}{\times}{\bf B}_M$. Therefore, we could
regard the vortex-antivortex pairs as point-like particle-antiparticle pairs 
coupling to the static e.m.-like Magnus field, $A_{M ~ \mu}$. The
spontaneous nucleation of these pairs in the bulk of the superconducting 
film is therefore a further cause of supercurrent decay and another important
source for the film's residual resistance. The decay rate is given by 
Eq. (\ref{rate0}) in our theory, in the absence of thermal fluctuations and
of a pinning potential. We notice that suggestions for a vortex-antivortex 
nucleation mechanism for supercurrent decay in superconducting films were
already put forth by Halperin and Nelson \cite{hane} in the context of a
temperature-driven Kosterlitz-Thouless unbinding of vortex-antivortex pairs. 
In our theory, the components of each pair eventually decouple (even at zero 
temperature and for any $ B_M$) 
due to quantum dissipation, which is the fundamental origin of 
the residual resistance \cite{ieju95}. The data of Ref. \cite{paro} can be
fitted by a law of the type $R=R_0\exp\{-a/(I-I_0)^b\}$, with $a$, $b$ and   
$I_0$ independent fitting parameters. This seems to agree with our formulas.

A much-researched way of hindering the (thermal- or quantum-) motion of the
vortices (thus reducing the resistance) is represented by the introduction of
pinning centers \cite{reviews}. In fact, 
periodic pinning would depress the vortices mobility.
However, the main focus of the present paper is rather on the 
vortex-antivortex nucleation. According to our results of Section VI,
the nucleation would be depressed for $A_0>0$ and favoured for
$A_0<0$, the latter case being probably the more relevant one as it
corresponds to the expectation that the presence of pinning centers 
could actually aid the nucleation. In any case, current-dependent
oscillations linked to the periodicity of the pinning lattice would
modulate the nucleation rate, see Eq. (\ref{fullrate}), 
and eventually could also appear in the induced resistance.
>From our estimates \cite{ieju96a} of the material parameters entering the
phenomenological description of the quantum nucleation we have predicted that
the oscillations ought to be observable. Superconducting films containing
lateral superlattice arrays of void or normal-metal holes have in fact begun 
to be fabricated and studied \cite{mosh}.

In conclusion, we have presented in a new light and context a systematic 
study of particle-antiparticle pair nucleation in the presence of static 
fields and of a dissipative medium. We have shown that pure CL quantum 
dissipation corresponds to the coupling to a space-like string, while full
relativistic space-time-like strings give rise to no vacuum decay.  We have 
applied our formalism to evaluate the pair-production rate induced by an 
e.m.-like field and demonstrated that in the presence of a background 
periodic potential oscillations connected to the underlying potential's 
periodicity will show up in the vacuum decay rate. Our theory could have 
interesting consequences for suitably-prepared superconducting thin films.

\vskip 2.0 truecm

\begin{center}
ACKNOWLEDGEMENTS 
\end{center}

\vskip 1.0 truecm 

R.I. is grateful to the Max-Planck Institut f\"ur Physik Komplexer Systeme, 
Au{\ss}enstelle Stuttgart, where this work was completed, for kind 
hospitality; G.J. is grateful to the International School for Advanced 
Studies (SISSA), Trieste, where part of this work was carried out, for 
hospitality. We are grateful to Dieter Weiss for pointing out Ref. 
\cite{mosh} to us. We acknowledge support from EC contracts No.s 
CHRX-CT920035, 
ERBFMRXCT96-0045 
and ERB4001GT957255.

\end{document}